\documentclass[twocolumn,nofootinbib,superscriptaddress]{revtex4-1}

\pdfoutput=1

\usepackage{slashed}

\usepackage[normalem]{ulem}

\usepackage{color}

\usepackage{epsfig}
\usepackage{epstopdf}
\usepackage{epsf}
\input{epsf.sty}
\usepackage{graphicx,amsmath,amssymb}
\usepackage{epsfig}
\allowdisplaybreaks

\def\ei{\end{itemize}}
\def\be{\begin{equation}}
\def\ee{\end{equation}}
\newcommand{\bea}{\begin{eqnarray}}
\newcommand{\eea}{\end{eqnarray}}

\usepackage{bbm,bm,amsmath,amssymb}

\usepackage{hyperref}

\def\K{{K\"{a}hler}}

\newcommand{\la}{\lambda}

\newcommand{\rf}[1]{(\ref{#1})}
\usepackage{float}

\begin{document}

\title{\Large dS Vacua and the Swampland}
\author{Renata Kallosh}
\email{kallosh@stanford.edu}
\affiliation{Stanford Institute for Theoretical Physics and Department of Physics, Stanford University, Stanford,
CA 94305, USA}
\author{Andrei Linde}
\email{alinde@stanford.edu}
\affiliation{Stanford Institute for Theoretical Physics and Department of Physics, Stanford University, Stanford,
CA 94305, USA}
\author{Evan McDonough}
\email{evan\_mcdonough@brown.edu}
\affiliation{Department of Physics, Brown University, Providence, RI, USA. 02903}
\author{Marco Scalisi}
\email{marco.scalisi@kuleuven.be}
\affiliation{Institute for Theoretical Physics, KU Leuven, Celestijnenlaan 200D, B-3001 Leuven, Belgium}

 \begin{abstract}
 
In this note we revisit some of the recent 10d and 4d arguments suggesting that the uplifting of supersymmetric AdS vacua leads to a flattening of the potential, preventing the formation of dS vacua. We explain why the corresponding 10d approach is inconclusive and requires  considerable modifications. We also show that while the flattening effects may occur for some extreme values of the parameters, they do not prevent the formation of dS vacua within the range of validity of the 4d KKLT models.  The KL version of the KKLT scenario based on a racetrack superpotential requires parametrically small uplifting,  which is not affected by flattening. We show that this scenario is compatible with the weak gravity conjecture for a broad choice of parameters of the KL model. Thus, the results of our analysis do not support  the recent swampland conjecture.

  \end{abstract}

\maketitle

\section{Introduction}

The string theory landscape scenario emerged in an attempt to solve two problems simultaneously: to provide a mechanism of de Sitter (dS) vacuum stabilization in string theory, and to solve the cosmological constant problem. One of the most popular versions of this theory is based on the KKLT scenario~\cite{Kachru:2003aw}, but other mechanisms of dS vacuum stabilization are also available. 

The most important part of dS constructions in string theory  is the enormous  multiplicity of vacuum states in the theory  and the possibility to tunnel from one of these states to another~\cite{Douglas:2003um,Douglas:2006es,Denef:2007pq,Kachru:2003aw,Susskind:2003kw}. This allows to use the anthropic constraint $|V_{dS}|\lesssim 10^{{-120}}$   to account for the incredible smallness of the vacuum energy:  If there were some vacua with the required energy density prior to the calculation of quantum corrections, then many  of them may move away from the anthropic range when quantum corrections are taken into account, but many other vacua will enter this range. Thus one should be able to find many suitable vacuum states   even though the precise vacuum energy in these states cannot be calculated with the  precision $|V_{dS}|\lesssim 10^{{-120}}$. 

This basic idea makes the string landscape scenario very robust. It cannot be invalidated by arguments of naturalness, by the weak gravity conjecture, 
  or by the possibility that radiative corrections may affect dS vacua. To disprove this scenario one would have to prove a no-go theorem, that would state that all of the $10^{{500}}$ or more dS vacua in string theory cannot exist, as suggested recently in the new swampland conjecture  \cite{Obied:2018sgi,Ooguri:2018wrx}, which we will call the no-dS conjecture. Despite many attempts to prove such no-go theorem  during the last 15 years, no such proof is available \cite{Akrami:2018ylq}. 
  
One could argue that even though the new swampland conjecture  \cite{Obied:2018sgi,Ooguri:2018wrx} may not have a derivation or proof, it can be validated if  future observations demonstrate that dark energy has equation of state different from the dS equation of state $w = -1$. However, this is not the case. First of all, all models of dark energy proposed in \cite{Obied:2018sgi} are already disfavored at a statistical significance higher than $4.5\,  \sigma$    \cite{Akrami:2018ylq,Raveri:2018ddi}.  Secondly, the models of dark energy with $w \not = -1$ can be constructed in the string theory landscape scenario with dS vacua. Moreover, the possibility to roll down towards the asymptotic dS state may simplify the construction of quintessence models, see e.g.   \cite{Akrami:2018ylq,Akrami:2017cir} and references therein.

Nevertheless, since this issue is extremely important,  one should carefully examine any piece of evidence against the string landscape scenario. One of the main arguments   in support of the no-dS conjecture was based on the suggestion of  \cite{Moritz:2017xto} that a certain version of the KKLT scenario does not lead to a consistent uplifting of the supersymmetric AdS vacuum to a metastable dS vacuum. 

The investigation performed in   \cite{Moritz:2017xto} consists of two parts: a detailed but very complicated 10d analysis, and a  4d investigation of the KKLT model, which is modified to take into account backreaction. 
The basic argument  of  \cite{Moritz:2017xto} is that, during the uplifting,  the potential flattens because of the backreaction to the 
uplifting, and dS minimum never forms.  The authors effectively propose that the backreaction to the uplifting is many orders of magnitude stronger that the basic effect of uplifting.  This is a very unconventional proposition, which should be carefully examined. 

Because of the complexity of the 10d   investigation, the authors of \cite{Moritz:2017xto}  were forced to make various assumptions, or rely upon results previously obtained  in  
different contexts.  In particular, some results of  the 10d investigation  of gaugino condensation in  \cite{Moritz:2017xto} contain divergences,  which required regularization and interpretation based on various conjectures  \cite{Moritz:2017xto}. The original analysis of these divergences in  \cite{Moritz:2017xto} contained an error corrected in \cite{Gautason:2018gln}.   Moreover, a recent  examination of this issue in   \cite{Hamada:2018qef,Kallosh:2019oxv} suggests that these divergences do not appear at all if one takes into account the 4-fermion non-derivative coupling, which was neglected in \cite{Moritz:2017xto,Gautason:2018gln}. We present a short discussion of this issue in section  \ref{10d}. This is just one of the many reasons to believe that  the uncertainties involved in the 10d investigation of \cite{Moritz:2017xto,Gautason:2018gln} make its results inconclusive \cite{Cicoli:2018kdo,Akrami:2018ylq,Kallosh:2018wme,Kallosh:2018psh}.

The 4d analysis of the KKLT model in  \cite{Moritz:2017xto} was supposed to confirm the 10d approach and the absence of dS vacua. However, a more detailed investigation in  \cite{Kallosh:2018wme,Kallosh:2018psh} demonstrated that the  modified 4d KKLT
model proposed in   \cite{Moritz:2017xto} is inconsistent, whereas all known consistent 4d versions of the KKLT scenario, in the domain of their validity, do not support the recent swampland conjecture \cite{Obied:2018sgi,Ooguri:2018wrx} and  are fully compatible with the existence of dS vacua in string theory.

The results of  \cite{Kallosh:2018wme,Kallosh:2018psh} are pretty straightforward and can be easily verified. However, recent comments on  these issues in \cite{Gautason:2018gln} show that a more detailed explanation of the relation between uplifting, flattening, and   dS vacua is warranted.   We  discuss these issues in sections \ref{basic} and
 \ref{flat}, and demonstrate by an explicit example that dS vacua do appear for proper choice of parameters even if one assumes extremely strong backreaction and flattening of the potential.

In section \ref{KL}  we  discuss the KL version of the KKLT mechanism \cite{Kallosh:2004yh}.  It is based on a particular version of the racetrack superpotential; it is free from the problems discussed in  \cite{Moritz:2017xto}.   Recently   Van Riet {\it et al} argued   \cite{Moritz:2018sui,Gautason:2018gln} that there might be some tension between this class of models and  a specific version of the  weak gravity conjecture introduced in  \cite{Moritz:2018sui}.  However,  the particular version of the weak gravity conjecture used in  \cite{Moritz:2018sui} is lacking evidence,\footnote{We are grateful to Westphal, McAllister and Rudelius for emphasizing this point to us.} and the authors provided no suggestions for how one could justify it.  Moreover,   an investigation of this issue by Blanco-Pillado {\it et al}  \cite{Blanco-Pillado:2018xyn} demonstrated that one can  satisfy even the unconventional version of the weak gravity conjecture introduced in  \cite{Moritz:2018sui} by a proper choice of parameters of the KL model. Our  investigation of this problem in section \ref{KL} confirms and strengthens  the results of \cite{Blanco-Pillado:2018xyn} for a broad set of parameters of the KL model. Thus, this model, which allows strong vacuum stabilization protected by supersymmetry \cite{BlancoPillado:2005fn,Kallosh:2014oja,Akrami:2018ylq}  and is most suitable for constructing realistic inflationary models \cite{Kallosh:2011qk,Dudas:2012wi,Kallosh:2018zsi}, does not seem to suffer from any problems with uplifting and stabilization of dS vacua mentioned in 
 \cite{Moritz:2017xto,Moritz:2018sui,Gautason:2018gln}.

\section{Gaugino condensation in 10${\rm \bf d}$} \label{10d}
 
Gaugino condensation refers to the formation of a non-vanishing fermion condensate Tr$\langle \la \la\rangle $ in the theory. It plays a significant role in the KKLT  construction of de Sitter vacua, since it provides a non-perturbative part of the superpotential $W_{np}= Ae^{-aT}$, where $T$ is the volume of the 4-cycle wrapped by condensing D7 branes. 
The fermionic part of the 10d analysis of \cite{Moritz:2017xto,Gautason:2018gln}  relies on earlier papers \cite{Marolf:2003ye,Camara:2004jj,Martucci:2005rb,Baumann:2010sx,Dymarsky:2010mf}, where gaugino condensation was studied at quadratic order in the fermions. Meanwhile, \cite{Moritz:2017xto,Gautason:2018gln} deduce from \cite{Marolf:2003ye,Camara:2004jj,Martucci:2005rb,Baumann:2010sx,Dymarsky:2010mf} the gaugino condensate contribution to the 4d effective action at \emph{quartic} order in the fermions, but neglect the intrinsic 4-gaugino interaction on D7-branes. This latter fact makes the analysis of \cite{Moritz:2017xto,Gautason:2018gln} incomplete, as pointed out in \cite{Hamada:2018qef,Kallosh:2019oxv}.

 Technically, the computation was performed   in \cite{Marolf:2003ye,Camara:2004jj,Martucci:2005rb,Baumann:2010sx,Dymarsky:2010mf} using the gauge-fixed $\kappa$-symmetric Dp-brane action. It was  explicitly stressed there that the result is only valid  in   the approximation quadratic in fermions.
 This action is supersymmetric and is known only in the case of a single Dp-brane with an Abelian vector multiplet. The relevant coupling was  presented    in \cite{Baumann:2010sx,Dymarsky:2010mf},
 \be
S_{D7}^{Maxw} \supset - {i\over 32\pi l_s^4} \int _{X^4}  d^4 x \bar \la_{\dot \alpha} \bar \la^{\dot \alpha} \int_D G_3 \cdot \Omega\,  J\wedge J +c.c.
\label{single} \ee
The D7-brane fills the four-dimensional space $X_4$  and wraps the four-cycle D  of the internal manifold. Note that here $G_3$  couples only to the fermion on a single D7, $\la$ does not have a non-Abelian gauge group index. 

The next step proposed in \cite{Camara:2004jj,Martucci:2005rb,Baumann:2010sx,Dymarsky:2010mf} is that for many coincident D7 branes, the  3-form couples to the trace of the fermion bilinears on the surface of $N$  coincident D7 branes, 
 \be
S_{D7}^{YM} \supset - {i\over 32\pi l_s^4} \int _{X^4}  d^4 x \mathrm{Tr} ( \bar \la_{\dot \alpha} \bar \la^{\dot \alpha} )\int_D G_3 \cdot \Omega J\wedge J +c.c.
\label{N} \ee
In \cite{Camara:2004jj,Martucci:2005rb,Baumann:2010sx,Dymarsky:2010mf} the issue of the 4-fermion coupling on the surface of the $N$ coincident D7 branes was not studied. Meanwhile, in  \cite{Moritz:2017xto} (see also \cite{Gautason:2018gln}) an  assumption was made  that replacing $\bar \la_{\dot \alpha} \bar \la^{\dot \alpha}$ by $ \mathrm{Tr} \, \bar \la_{\dot \alpha} \bar \la^{\dot \alpha} $ when describing many D7 branes is the only generalization required. The authors of \cite{Moritz:2017xto}  evaluated  the corresponding 4-fermion terms  using $G_3$ equations of motion, and found divergences, which required regularization.

The issues with the 4-fermion 10d analysis in  \cite{Moritz:2017xto,Gautason:2018gln} were recently discussed in \cite{Hamada:2018qef,Kallosh:2019oxv}. In  \cite{Hamada:2018qef} a  proposal was made that the 4-fermion terms should be present on the D7 brane,  which  might resolve the disagreement with the 4d analysis. They suggested to follow the related set up in 
 \cite{Horava:1995qa,Horava:1996ma,Horava:1996vs} for an M-theory compactified on a one-dimensional interval.  The action in  \cite{Horava:1995qa,Horava:1996ma,Horava:1996vs} has a perfect square term
 \be
 - \int d^{11} x \sqrt g \Big (G_{\mu\nu\rho11} - c\, \delta(x_{11})  \mathrm{Tr}\la \Gamma_{\mu\nu\rho} \la\Big )^2  ,
 \ee
and the $\delta(0)$ singularities cancel due to the perfect square structure in the action, which requires the presence of the 4-fermion term. A proposal made in \cite{Hamada:2018qef}  for type IIB theory with coincident compactified D7 branes
has additional features, which, in particular, allow to reach a nice correspondence with the 4-fermion terms in 4d action.

 In \cite{Kallosh:2019oxv} it was stressed that, as shown in \cite{Ferrara:1982qs,Cremmer:1982en}, the  4d supergravity action has a term with the  square of the auxiliary field,  $|F|^2$. And since the on shell value of the auxiliary field   involves gaugino 
 \be
F^\alpha = - e^{K/2} g^{\alpha \beta} \overline \nabla_{\bar \beta} \overline W     + {1\over 4} \bar f_{AB\bar \beta} g^{\bar \beta \alpha} \bar \lambda^A P_L   \lambda^B\, , 
\ee
 one finds that the bilinear and quartic dependence on gaugino's comes via a perfect square in the action, $|F|^2$.

It was also pointed out in  \cite{Kallosh:2019oxv} that the existence of the perfect square term in
10d action is a general feature of the  Einstein-Yang-Mills supergravity \cite{Bergshoeff:1981um,Chapline:1982ww,Dine:1985rz} and of the underlying  10d superspace geometry. The complete 10d supergravity action, where we suppress terms depending on gravitino and dilatino but keep all terms with gaugino and bosons, is
 \bea\label{EYMaction}
&e^{-1} {\cal L} &\, =  - {1\over 2}  R - {3\over 4}\phi^{-{3\over 2}} \Big ( F_{\mu\nu\rho}^{YM}   - {\sqrt 2\over 24} \, \phi^{3\over 4} \, \mathrm{Tr} \,  \bar \la \Gamma_{\mu\nu\rho} \la\Big )^2  \nonumber \\ 
&&-  {9\over 16} \Big ({\partial_\mu \phi\over \phi}\Big )^2 -{1\over 4}  \mathrm{Tr} \, (F_{\mu\nu}^{YM})^2 -{1\over 2} \mathrm{Tr} \,  \bar \la \gamma^\mu D_\mu \la \, .
\eea
Type IIB theory in presence of calibrated Dp-branes and Oq-planes as local sources has only 1/2 of the maximal 10d supersymmetry. Therefore the effective action describing IIB supergravity with coincident D9-branes is an action of the Einstein-Yang-Mills supergravity, up to the Born-Infeld type terms with higher derivatives. Therefore  gauginos which live on the coincident D9 branes must have a 4-fermion coupling. Dimensional reduction to coincident D7 branes suggests that they also must have a 4-fermion coupling.

The reason why this was not discussed in \cite{Marolf:2003ye,Camara:2004jj,Martucci:2005rb,Baumann:2010sx,Dymarsky:2010mf} is that their derivation of eq. \rf{single}  was based on the $\kappa$-symmetric Abelian D7-brane action, which is supersymmetric upon gauge-fixing. For the Abelian vector multiplet which lives on the brane the 4-fermion gaugino interaction vanishes since $ ( \bar \la \Gamma_{\mu\nu\rho} \la )^2=0$. Therefore, even if in \cite{Marolf:2003ye,Camara:2004jj,Martucci:2005rb,Baumann:2010sx,Dymarsky:2010mf} the issue of the non-derivative 4-fermion coupling would be raised, the answer would be negative, based on the $\kappa$-symmetric Abelian D7-brane action.  

But in \cite{Moritz:2017xto,Gautason:2018gln}  the dependence on non-Abelian gaugino, derived in \cite{Marolf:2003ye,Camara:2004jj,Martucci:2005rb,Baumann:2010sx,Dymarsky:2010mf} and shown here in eq. \rf{N}, was assumed to be valid without any additional  gaugino-dependent terms even for non-Abelian vector multiplets on D7.
From  Einstein-Yang-Mills supergravity \rf{EYMaction} it follows that the bilinear in gaugino terms must be accompanied by the quartic gaugino term. This means that the assumption in \cite{Moritz:2018sui,Gautason:2018gln}  about the absence of 4-fermion terms in the action \rf{N} cannot be valid, and therefore the whole 10d analysis needs to be reconsidered. It is likely that it will reduce to the construction of the kind given in \cite{Hamada:2018qef}, and the consistency with 4d physics in this aspect will be restored.
In what follows we will concentrate on the 4d analysis of the KKLT construction.

\section{KKLT model and its consistent generalizations}\label{basic}
The KKLT model in the 4d supergravity formulation can be described by a superpotential 
\be\label{WKKLT}
W= W_0 +Ae^{-aT} +  bS .
\ee
Here $S$ in the nilpotent multiplet representing  anti-D3 brane responsible for the uplifting \cite{Ferrara:2014kva,Kallosh:2014wsa,Bergshoeff:2015jxa}. 
The \K\, potential can be either 
\be K=-3\log\left(T+\bar T \right) +{S \bar S}, 
\ee
or
\be\label{KKK}
K=-3\log\left(T+\bar T -{S \bar S}\right) \ .
\ee
The modification proposed in \cite{Moritz:2017xto}
introduces an extra term $ c Ae^{-aT} S$ in the superpotential, with an extra parameter $c$  describing effects of backreaction\, 
\be\label{W1}
W= W_0 +Ae^{-aT} + c Ae^{-aT} S+ bS \ .
\ee
It was argued in \cite{Moritz:2017xto} that $|c A | \gg b$.  This  would imply that the backreaction  is much greater than the main effect. This is an unusual proposition which does not seem  well motivated  \cite{Kallosh:2018wme, Cicoli:2018kdo}. 

 Instead of debating the reliability of the assumption $|c A | \gg b$, we studied the general case, including $|c A | \ll b$ as well as $|c A | \gg b$.  According to  \cite{Moritz:2017xto}, the value of $c$ is exponentially sensitive to various  parameters of compactification, so in the string theory landscape it may take many different values for any given $b$. Therefore, following  \cite{Kallosh:2018wme,Kallosh:2018psh}, we will analyze the potential as a  function of two independent parameters $b$ and $c$, and check whether it may have dS vacua. 

After \K\, transformation $(c Ae^{-aT} +b) S \rightarrow  S$ the model can be equivalently represented as 
\be\label{wnob}
W= W_0 +Ae^{-aT} + S ,
\ee
\be\label{wv1}
K=-3\log\left(T+\bar T -{S {\bar S}
\over  |c Ae^{-aT} +b|^2}\right) .
\ee
The denominator in \rf{wv1} is a perfect square, which is positive everywhere except the point
$
T_0= \frac{1}{a}\ln (-cA/b )
$,
where it vanishes. As emphasized in \cite{Kallosh:2018wme,Kallosh:2018psh}, this makes the use of the nilpotent multiplet $S$ in the  model \rf{W1} inconsistent for $|c A | > b$.  (A similar problem appears at large $c$ in the model proposed in \cite{Moritz:2018ani}; see \cite{Kallosh:2018psh}). The inconsistency disappears for $|c A | < b$, because in that case the point $
T_0= \frac{1}{a}\ln (-cA/b )
$ is outside the physical domain with  $\rm Re \, T >0$. In that case,  the effects of backreaction are exponentially suppressed, and dS uplifting  occurs as in the original model \rf{WKKLT}.

Note that the inconsistency of the 4d model proposed in \cite{Moritz:2017xto} is rather subtle; it manifests itself only if one considers the full supergravity model including fermions. If one temporarily ignores this issue and calculates the bosonic potential,
one finds that the theory does contain a large family of dS vacua \cite{Kallosh:2018wme}.

To avoid the inconsistency  problem altogether, one can add a positive number or function to $ |c Ae^{-aT} +b|^2$ in \rf{wv1}. This removes the pole in the denominator  in \rf{wv1}, which makes the theory consistent. For example, one may consider the \K\ potential 
\be\label{wv3}
K=-3\log\left(T+\bar T -{S {\bar S}
\over  |c Ae^{-aT} +b|^2+\beta \,c^2 A^2  e^{-a(T+\bar T)}}\right) \,
\ee
where $\beta$ is some positive number \cite{Kallosh:2018psh}. This immediately makes $K^{S\bar S}$ strictly positive definite, which avoids all inconsistencies of the models of  \cite{Moritz:2017xto,Moritz:2018ani} for any choice of $\beta > 0$.  

\begin{figure}[h!]
\begin{center}
\includegraphics[width=7.7cm]{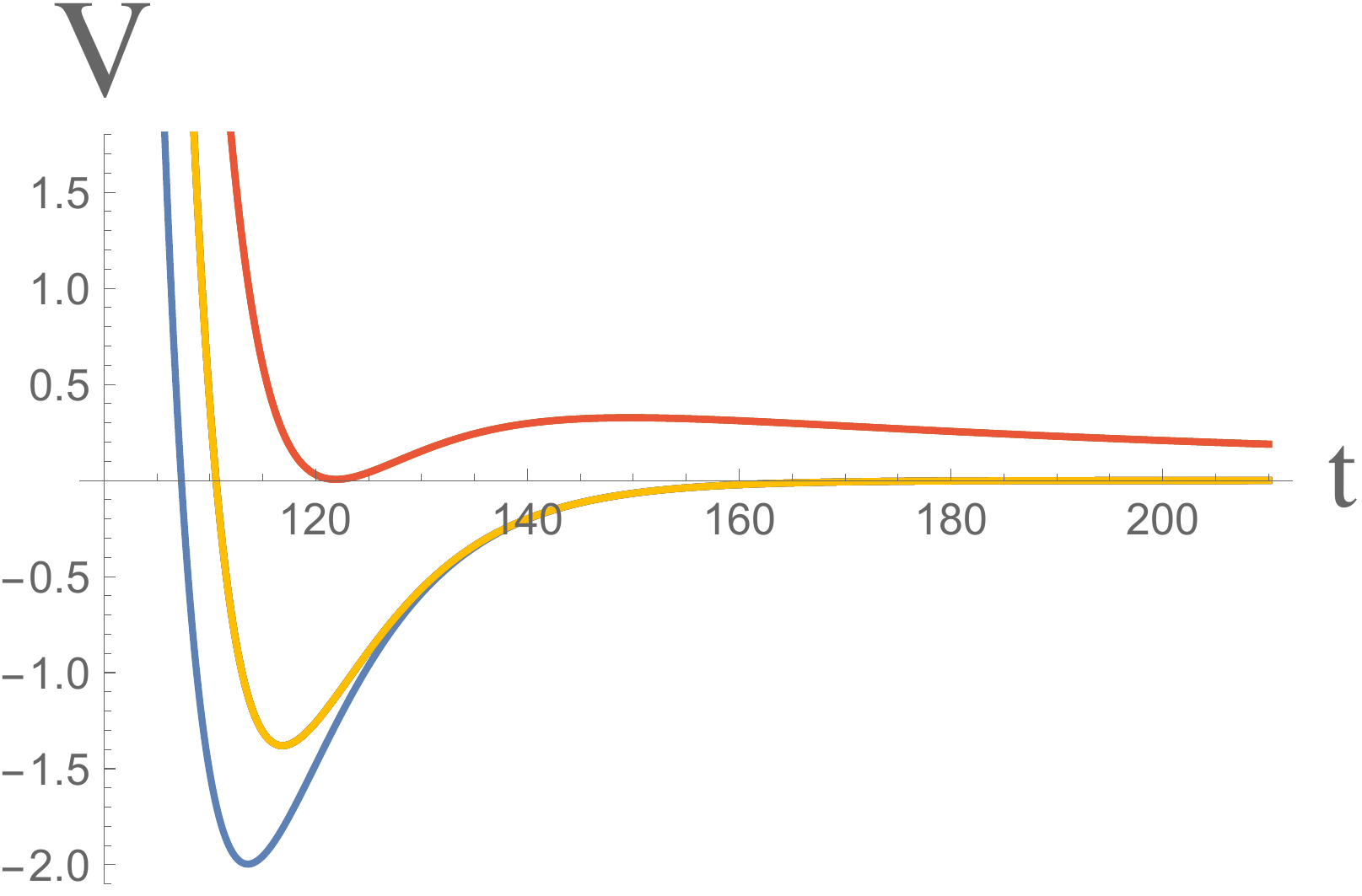}
\caption{The potential of the theory \rf{wv3} (multiplied by $10^{15}$) for  $A=1$, $a=0.1$, $W_0=-10^{-4}$,   $0<\beta\ll 1$. The blue (lower) line shows the potential with a supersymmetric AdS minimum prior to uplifting, at $b = c = 0$. The second (yellow),  line shows the potential at $b = 0$ uplifted by increase of $c$ to  $c = 1$.  Finally, the upper (red) line shows the potential with a dS (nearly Minkowski) minimum for $c = 1$, $b =10^{{-5}}$. The main part of the uplifting is not due to the large change of $c$ from 0 to 1, but due to the tiny change of $b$ from 0 to $b =10^{{-5}}$.}
\label{f2}
\end{center}
\vspace{0cm}
\end{figure}

For $0<\beta\ll 1$, the KKLT potential in this model practically coincides with the potential obtained in our paper  \cite{Kallosh:2018wme}. Thus our results about the existence of dS vacua contained in  \cite{Kallosh:2018wme} are confirmed for a large range of parameters for consistent generalizations of the KKLT scenario \cite{Kallosh:2018psh}. In particular, for small $\beta$ 
the potential in the model \rf{wv3} practically coincides with the potential shown in one of the figures in   \cite{Kallosh:2018wme}, which we reproduce here for convenience. The potential is shown as a function of $t = {\rm Re}\, T $.

Yet another consistent model  is described by
 $W$ \rf{wnob} and \K\ potential 
\be\label{wv4}
K=-3\log\left(T+\bar T -{S {\bar S}
\over  |b|^2+ |c|^2 e^{-a(T+\bar T)}}\right), \,
\ee
This model is consistent for any choice of $\{b,c\}$. It also has  dS vacua for a broad choice of its parameters  \cite{Kallosh:2018psh}.

\section{Flattening effects in KKLT}\label{flat}

Recently there was a new paper on this subject, by Van Riet {\it et al} \cite{Gautason:2018gln},  which commented on the 10d and 4d analysis of the  KKLT scenario in  \cite{Moritz:2017xto,Kallosh:2018wme, Moritz:2018ani,Kallosh:2018psh}. 
 
With respect to 10d, the authors  of \cite{Gautason:2018gln} emphasized a problematic issue  related to the  divergences which appear  in the calculations of  \cite{Moritz:2017xto}.  However, as we already discussed in section \ref{10d}, investigation of this issue in  \cite{Moritz:2017xto,Gautason:2018gln} missed important terms  \cite{Hamada:2018qef,Kallosh:2019oxv}. 
Thus, in agreement with \cite{Kallosh:2018psh,Cicoli:2018kdo}, we believe that the 10d analysis in  \cite{Moritz:2017xto, Gautason:2018gln} is inconclusive. Moreover, according to  \cite{Moritz:2017xto},  their 10d analysis, as well as the one in  \cite{Gautason:2018gln}, is model-dependent; it does not apply at all to the version of the KKLT scenario where the moduli stabilization is provided not by gluino condensate but by the Euclidean D-brane instanton effects. 

Therefore in the present paper, following \cite{Kallosh:2018psh}, we  concentrate on the 4d KKLT models.  The comments on these models given in  \cite{Gautason:2018gln} are very short and somewhat confusing. We feel that  some clarifications are in order.

Ref.  \cite{Gautason:2018gln} makes a conjecture that if  the parameter $c$ should be suppressed in order to have a well-defined nilpotent description, then the nilpotent description is not adequate to model the flattening effects, which was the main focus of investigation in    \cite{Moritz:2017xto,Gautason:2018gln}.

This conjecture consists of two incorrect parts. In the present paper we constructed several consistent 4d generalizations of the original KKLT model. These models remain valid for any value of $c$, and they adequately describe the flattening effects, which indeed take place in these models as expected in \cite{Moritz:2017xto}. Thus there are consistent models where the constant $c$ can be large, and the flattening effect is real. However, as we are going to explain, flattening  {\it does not} imply the absence of dS vacua, contrary to what was conjectured and many times repeated in  \cite{Moritz:2017xto,Gautason:2018gln}.   dS vacua   can be found in all of these models, even  for extremely large $c$, despite the flattening of the potential.

Strictly speaking, one should not study  these issues in the  model of \cite{Moritz:2017xto}, as the authors of  \cite{Gautason:2018gln} do, since this model is inconsistent at large $c$. However, as we already mentioned in the previous section,  the fully consistent model \rf{wv3} constructed in our paper  \cite{Kallosh:2018psh} leads to the same bosonic potential as in  \cite{Moritz:2017xto},  in the small $\beta$ limit. In this sense, investigation of the bosonic potential of  \cite{Moritz:2017xto}   provides    correct information about the uplifting in the consistent  model \rf{wv3} at small $\beta$.

\begin{figure}[!ht]
\begin{center}
\includegraphics[width=7.8cm]{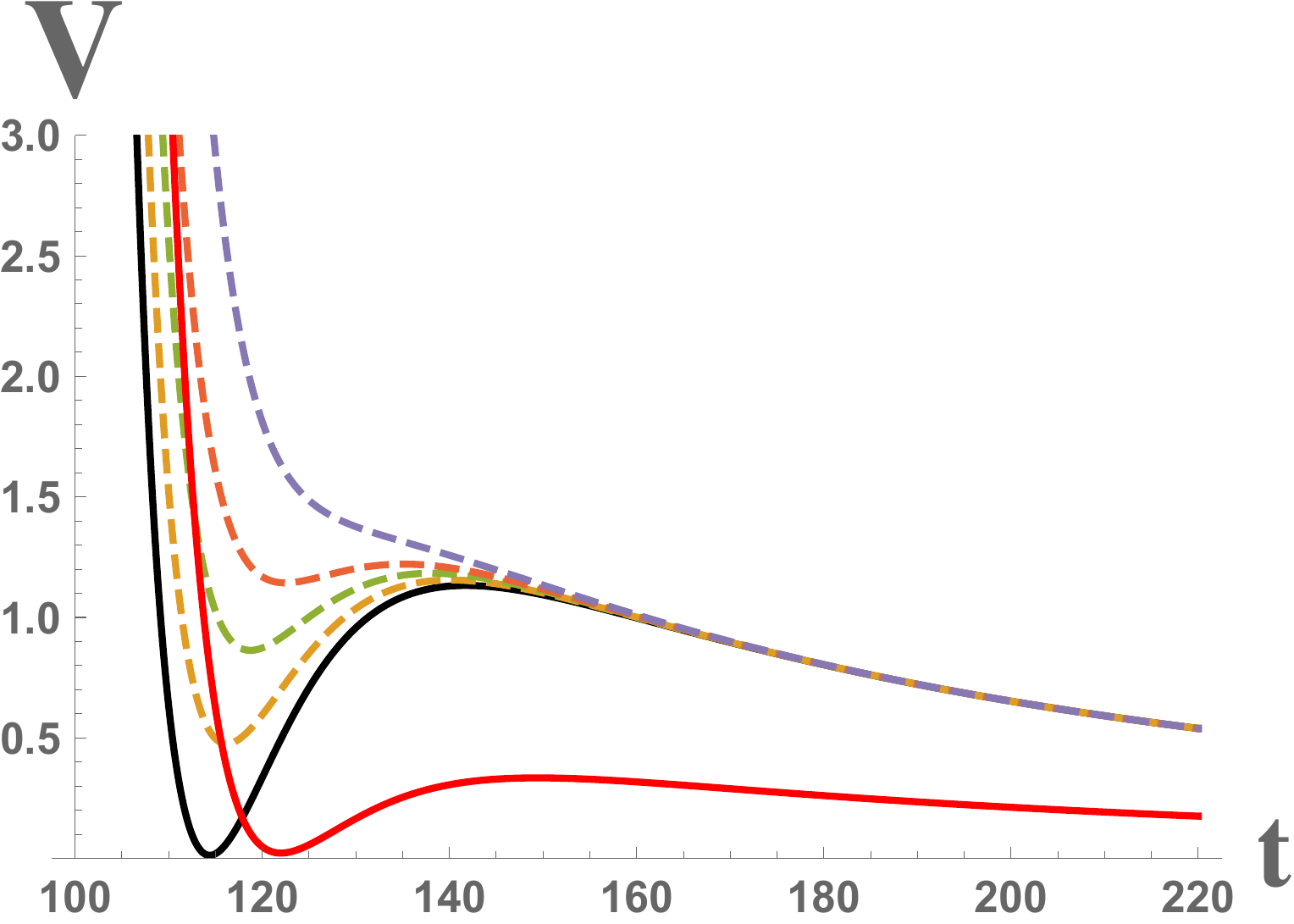}
\caption{The potential in the model \rf{W1} (multiplied by $10^{15}$) for  $A=1$, $a=0.1$, $W_0=-10^{-4}$. The black   line shows the potential at $c = 0$ uplifted by increase of $b$ to from $0$ to $b = 1.77 \times 10^{{-5}}$. The dotted lines show further uplift due to increase of $c$ from $0$ to $c = 0.2,\   0.4, \ 0.6$, and $1$. The last increase of $c$ leads to  destabilization of the dS minimum.  The red line shows that if one takes $c = 1$ and  decreases the value of $b$ from  $b = 1.77 \times 10^{{-5}}$ to  $b = 10^{{-5}}$, one finds a stable dS minimum, the same that is shown by the red line in Fig. \ref{f2}, and in Fig. 2 in \cite{Kallosh:2018wme}.}
\label{stab}
\end{center}
\end{figure}

 However, while it indeed  makes sense to study uplifting in the model of  \cite{Moritz:2017xto},   the authors of  \cite{Gautason:2018gln}   do something  different:  They instead study destabilization of  dS vacua, which occurs {\it after} the uplifting, during the further increase of  $c$ and $b$. The discussion was based on the Fig. 2 given in   \cite{Gautason:2018gln}, which we reproduce here, with some modifications, using the original notations for the model \rf{W1}, see Fig. \ref{stab}. 
 
The black solid line in Fig.~\ref{stab} shows the dS  vacuum with a small positive vacuum energy obtained by uplifting with $c = 0$, $b = 1.77 \times 10^{{-5}}$. The dotted lines show further dS uplift due to increase of $c$ from $0$ to $c = 0.2$, $c = 0.4$, $c = 0.6$ for $b = 1.77 \times 10^{{-5}}$. As we see, the last increase of $c$ leads to  destabilization of the dS minimum. 

This clearly shows that there is a large family of dS vacua in this theory, even for $c$ many orders of magnitude greater than $b$. These results give a simple 4d counter-example for the statements of  \cite{Moritz:2017xto,Gautason:2018gln}. Instead  of coming to this obvious conclusion, the authors of  \cite{Gautason:2018gln} argue that this scenario is problematic because if one continues increasing $c$ up to $c = 1$, this leads to destabilization of the uplifted dS vacua.

Indeed, it was shown long ago that an excessively strong uplifting leads to dS destabilization \cite{Kallosh:2004yh}. This effect occurs even for $c=0$, if  $b$ is too large, so one should choose the uplift parameters carefully \cite{Kallosh:2004yh}. However, this  effect has no relation whatsoever to the possibility to uplift from AdS to dS. 
  It is very easy to cure the destabilization of the dS minimum at $c = 1$, $b = 1.77 \times 10^{{-5}}$. It is sufficient to take a slightly smaller value  of $b = 10^{{-5}}$, which leads to a stable dS minimum shown by the red line in Fig. \ref{stab}. It is the same red line that is shown in Fig. 1 in the previous section.  

The lesson that follows from this calculation is that the increase of $c$ from 0 to 1 practically does not affect the uplift: The effect of this increase is completely compensated by the less than 50\% decrease of $b$ from  $1.77 \times 10^{{-5}}$ to $10^{{-5}}$. Moreover, as  argued in  \cite{Kallosh:2018psh}, there is not much reason to even consider the regime with $c \gg  b$, because it would imply a physically  unreasonable situation where the backreaction to the small impact provided by $b$ is many orders of magnitude stronger than the original impact.  We considered the large $c$ regime only to show that even in this case the consistent dS uplift is possible.

\begin{figure}[!h]
\begin{center}
{ \includegraphics[width=7.8cm]{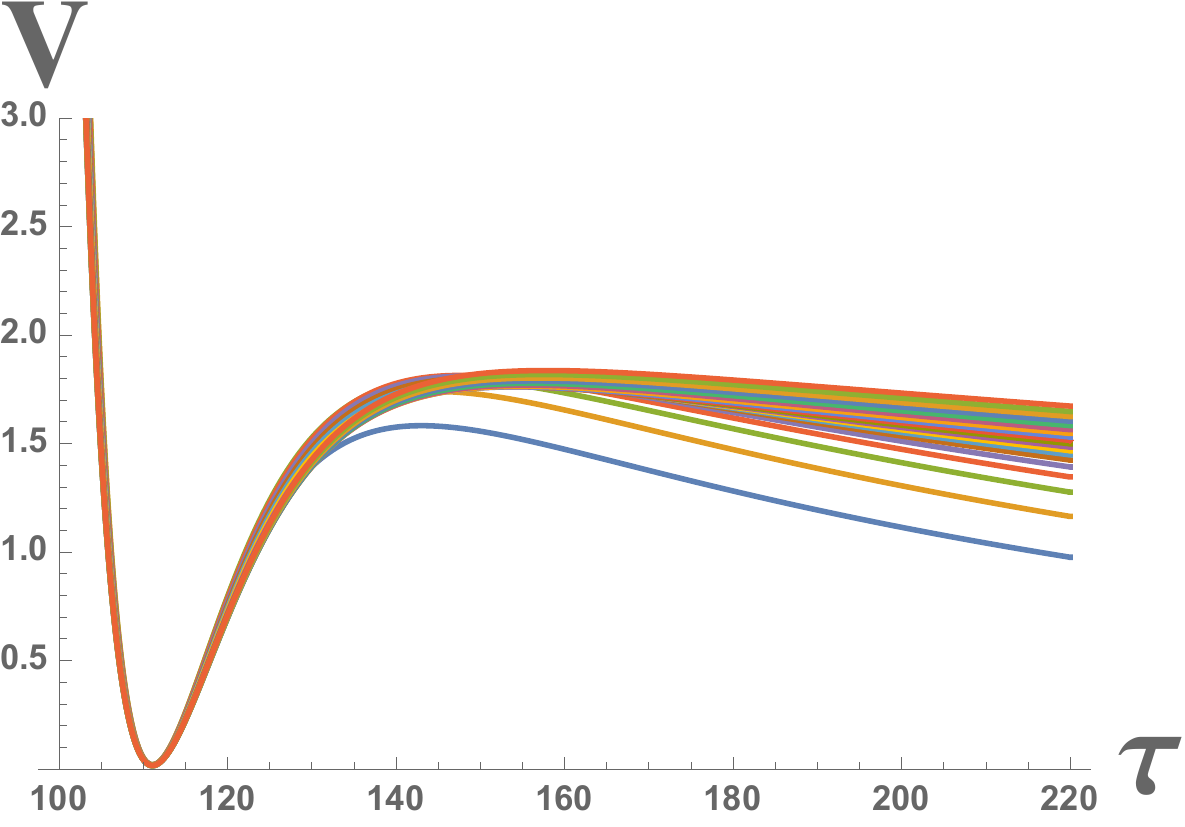}}
\caption{The potential in the model \rf{W1} multiplied  by $10^{15} c^{2.02 } \log c$ for   $t = \tau +  20 \log c$, $b c \approx 1.5 \times 10^{{-5}}$ and $c$ in the broad range from $c = 10^{2}$ to $c = 10^{20}$. }
\label{f5}
\end{center}
\end{figure}

But what if $c  = \mathcal{O}(1)$, which is 5 orders of magnitude greater than the required value of $b$,  still does not seem large enough to convince some authors? To address this question, we  now consider  an incredibly broad range of values of $c \gg 1$, all the way up to $c = 10^{20}$, for $A=1$, $a=0.1$, $W_0=-10^{-4}$, see Fig. \ref{f5}. To present our results for such a large range of  $c \gg 1$ in a single figure, we   rescale  the potential by multiplying it by $10^{15} c^{2.02} \log c$, and  plot  it as a function of  $\tau = t - 20 \log c$. We   find  a set of dS minima for each  value of $c \gg 1$, for $b c \approx 1.5 \times 10^{{-5}}$.  The minima of all of these potentials are at the same point $\tau = t - 20 \log c \approx 112$,  with the same curvature at the minimum, and approximately the same height of the barrier, making the dS vacuum metastable.

One may wonder how can one see the promised flattening effect at large $c$ in Fig. \ref{f5}? The answer is that the  height of the barrier in this figure was enhanced by the factor $\sim c^{2}$ in order to reveal the similarity of the shape of the potential for various $c$. Thus the true height of the barrier decreases approximately as $c^{{-2}}$ for large $c$, so the flattening effect is very real,  even though it occurs only for unreasonably large $c$. But we also see that this effect {\it does not} forbid dS vacua.

 The opposite statement made in \cite{Moritz:2017xto,Moritz:2018ani,Gautason:2018gln} was based on an appealing but incorrect  conjecture  that one can ignore $b$  if $c \gg b$. This conjecture was not based on any actual investigation. It is incorrect, because the term proportional to $c$ in \rf{W1} is exponentially suppressed at large $T = t$. That is why the position of the minimum at large $c$ shifts towards greater values of $t = \tau +  20 \log c$, see Fig. \ref{f5}. Thus, our results do not reveal any relation between the flattening effect and the existence of dS vacua. In particular, Fig. \ref{f5} demonstrates that well-formed metastable dS minima may exist even if one assumes that $b$ is 45 orders of magnitude smaller than $c$, and the potential is flattened by more than 40 orders of magnitude.

\section{The KL model and the Weak Gravity Conjecture}\label{KL}

A year after the  invention of the KKLT model~\cite{Kachru:2003aw}, it was recognized that combining this model with inflation effectively leads to an additional contribution to $\mu^{2}$, which could destabilize the volume modulus in the very early universe~\cite{Kallosh:2004yh}. The destabilization may occur at a large Hubble constant because the height of the barrier in the KKLT scenario is proportional to the square of $W_{0}$ related to the gravitino mass and the strength of supersymmetry breaking, which was often considered small. 

This problem disappears if supersymmetry breaking in this theory is sufficiently high, but there are several other ways to stabilize the KKLT potential. The simplest one, proposed in~\cite{Kallosh:2004yh}, is to change the superpotential to the racetrack potential with two exponents,  
\be
W_\text{KL}(T,S)  =W_{0} +Ae^{-aT}- Be^{-bT}  +   \mu^{2} S\ ,
\label{adssup}
\ee
 which can arise, for example, in the presence of two stacks of D7 branes wrapping homologous 4-cycles. 
Gaugino condensation on the first one is responsible for the KKLT-type term $Ae^{-aT}$, the second one for the term $- Be^{-bT}$. If there are $N_{1}$ branes in the first stack, and $N_{2}$ branes in the second one, one has $a = {2\pi/  N_{1}}$ and $b = {2\pi/  N_{2}}$. We assume, without loss of generality, that $a > b$. The parameters $A$ and $B$  depend on  the  values at which the complex structure moduli are stabilized \cite{Burgess:1998jh,Baumann:2006th,Baumann:2010sx}, and therefore one may expect $A$ and $B$ to span large range of possible values, 
due to the  large variety of vacua in the string theory landscape.

In what follows we will consider the models  where $a, b, A, B > 0$,   and 
\be\label{w0stab}
W_0=  -A \left({a\,A\over
b\,B}\right)^{a\over b-a} +B \left ({a\,A\over b\,B}\right) ^{b\over b-a}  \ .
\ee
For $\mu = 0$, the potential $V(T)$ has a stable supersymmetric Minkowski minimum at $T = t_{0}$, which can be found by solving two equations, 
\be\label{susy} 
W(t_{0})=0 \ , \qquad DW(t_{0})=0 \ ,
\ee
which yields
\be
t_{0}= {1\over a-b}\ln  {a\,A\over b\,B} \, .
\label{sigmacr} \ee

Adding a small correction to $W_{0}$ makes this minimum AdS. For $\mu \neq 0$, this minimum  can be easily uplifted to dS while remaining strongly stabilized \cite{Kallosh:2004yh,BlancoPillado:2005fn,Kallosh:2014oja}. Importantly, the height of the barrier in this scenario is not related to supersymmetry breaking and can be arbitrarily high. Therefore, this  potential can be strongly stabilized by a proper choice of the parameters, which makes it  especially suitable for being a part of the inflationary theory~\cite{Kallosh:2011qk,Dudas:2012wi,Kallosh:2018zsi}.

The KL model \rf{adssup} with the racetrack superpotential is just one of the many possible ways to find nearly supersymmetric vacua. One may consider general superpotentials which may emerge among $10^{500}$ or more versions of string theory compactification, after all quantum corrections are taken into account. 
If there is a point in the moduli space where supersymmetry is unbroken in Minkowski space (e.g. where equations \rf{susy} have a solution), then the tiny uplift required for describing dS space with an incredibly small cosmological  constant $\sim 10^{{-120}}$ does not lead to vacuum destabilization \cite{BlancoPillado:2005fn,Kallosh:2014oja}. We should emphasize that for our purposes (the theory of vacuum stabilization)  we do not need to require flatness of the potential, large excursions of the field away from the vacuum state, etc. We only need to make sure that the depth of the AdS (nearly Minkowski) minimum is parametrically smaller than the squares of the masses of all moduli in the vicinity of the minimum. This can be achieved in the KL construction.

If the number of possible vacua in the landscape is large enough, the number of nearly supersymmetric vacua should be also extremely large, which should address the general problem of uplifting in such vacua.  However, it is always nice to have explicit examples of the models where it may happen in a controllable way.

The status of such models  can be probed by applying simple consistency tests. Just like in the original version of the KKLT scenario, one should check that the radius of compactification $t_{0} $ sufficiently large, which is necessary for consistency of the supergravity interpretation. One can achieve it, for any $A$ and $B$, by considering sufficiently small  $a$ and $b$. 

In analogy with axion scalar potential generated by instantons, and with the knowledge that D-brane instantons enter the theory via the superpotential \cite{Witten:1996bn,Harvey:1999as}, one might anticipate higher order non-perturbative contributions to the gaugino condensate superpotentials in the KL model. The precise sense in which this intuition from instantons can be applied to gaugino condensation is via the dual description of gaugino condensation as Euclidean M5 brane instantons in M-theory (see e.g.~\cite{Denef:2008wq}  for a detailed technical  discussion of related issues).  Putting aside the technical details of the origin of such terms, one would expect that  higher order non-perturbative contributions to the gaugino condensate superpotential should be under control near the minimum of the KL potential if  
\be\label{large}
e^{-a \, t_{0}} < 1\, , \qquad  e^{-b \, t_{0}} < 1 \, .
\ee
Note  that the terms $e^{-a \, t_{0}}$ and $e^{-b \, t_{0}}$   are  strongly suppressed even for moderately large values of $a \, t_{0}$ and $b \, t_{0}$,
\be\label{linear}
a \, t_{0} > 1\, , \qquad b \, t_{0} > 1 \ .
\ee

Using \eqref{sigmacr}, we can represent the  conditions \eqref{large} as follows:
\be\label{KLWGC}
e^{-a \, t_{0}}= \left({b\,B \over a\,A}\right)^{a\over a-b}  <  1\, , \qquad e^{-b \, t_{0}} = \left({b\,B \over a\,A}\right)^{b\over a-b} <  1\, \ .
\ee
 All these conditions are satisfied, for example, if    $b\gtrsim a/2$ and   $A \gtrsim 2B$.  Thus the KL model is  expected to be consistent and reliable in the vicinity of the minimum of its potential $t_0$ for a broad choice of parameters.

Consider, for example, the case  $b = a/2$. Then one can check that the strongest constraint originates from the second inequality,\be\label{KLWGC2}
 e^{-b \, t_{0}} =  { B \over 2\,A}  <  1\, \ .
\ee
 This expansion parameter is smaller than 1/4 for $A > 2B$, and is smaller than  $10^{{-1}}$ for $A > 5B$.    It becomes even easier to satisfy all conditions mentioned above for $a-b \ll a$.

Then where does the statement of Moritz and Van Riet   \cite{Moritz:2018sui} that the KL model may contradict the  weak gravity conjecture  come from? 

To understand it, let us discuss the relation between the  conditions  of the type of   \rf{large}, \rf{KLWGC} and the weak gravity conjecture for the simplest axion inflation models with the Higgs-type potential $\sim (|\phi|^{2}-f_{a}^{2})^{2}$, as in natural inflation.  For such axions, there is a series of instantons with action  $S_{n} = O(n S_{1})$, where $n = 1, 2, ...$, see e.g.  \cite{Blumenhagen:2018hsh}. Here $f_{a}$ is the axion decay constant, which describes, up to a factor   $2\pi$, the periodicity of the axion potential. One can trust the axion potential ignoring higher order instanton corrections $\sim  e^{- n S_{1}}$ in the natural inflation scenario only if $e^{- S_{1}} < 1$, i.e. if $f_{a} < 1$. This is the condition that makes natural inflation so difficult to implement.  

 On the other hand, if, like in the string theory landscape, we do not care about natural inflation,  about quantum corrections and the shape of the potential in the angular direction, and only want to make sure that the potential has some minimum somewhere, then the constraint   $e^{- S_{1}} < 1$  becomes less relevant.  
 
  Now let us return for a moment to  the KKLT scenario with a superpotential with a single exponent  $W   =W_{0}- A e^{-  aT}+b S$ \rf{WKKLT}. The requirement of the exponentially  large suppression of  nonperturbative effects in a vicinity of the minimum of the KKLT potential  requires the condition $e^{-a \, t_{0}} < 1$, with $t_0$ being the position of the minimum. Just as in the natural inflation scenario, this requirement is equivalent, up to a factor $\sqrt {3/2}$, to the weak gravity conjecture requirement that the effective axion decay constant $f_{a} \lesssim 1$, where $f_{a}$ describes the periodicity of the KKLT potential in the ${\rm Im}\, T$ direction in canonical variables $\theta = \sqrt{2\over 3} {\alpha\over t_{0}}$  for  ${\rm Re}\, T = t_{0}$ \cite{Moritz:2018sui}.  Thus, just as in the usual axion scenario, the smallness of the effective axion decay constant $f_{a}$ in the KKLT model appears as a consequence of the requirement that the higher order nonperturbative effects are suppressed, $e^{-a \, t_{0}} < 1$.  

However, already at this level the situation is somewhat different from the one  in natural inflation. First of all, we are unaware of any higher order KKLT instantons with $S_{n} = O(n/f_{a})$ for $n >1$. Therefore, while the requirement that higher order nonperturbative corrections should be suppressed seems reasonable, at this stage  it  is somewhat  speculative. Secondly, there is no inflation in the axionic direction in the KKLT model, so we do not need to know the exact form of the KKLT potential in the axion direction. As we already discussed, the main requirement is that it should have a dS minimum {\it somewhere}, after potentially significant quantum corrections are taken into account.  However, it is important to have $t_{0} \gg 1$, to ensure that the supergravity approach to string theory is adequate. To be on the safe side, one may also impose the condition $ a \, t_{0}  > 1$.  One can easily find parameters which satisfy both of these conditions, but  in general, the condition $a \, t_{0}> 1$ may or may not be  required for the existence of  dS  vacua in KKLT. 

 Note that the most important features of the KKLT potential, such as the position of the AdS minimum, the position of the dS minimum after uplifting,  and the potential barrier responsible for the vacuum stabilization  are concentrated at $t \sim t_{0}$. Therefore all consistency constraints mentioned above are imposed on the theory at $t \sim t_{0}$.  The theory may become unreliable for $t \ll t_{0}$, and indeed we do not expect that the supergravity approximation is adequate for $t \lesssim 1$, but this is not our immediate concern as long as $t_{0} \gg 1$. 

Finally, we return to the KL scenario. The full potential as a function of $T = t + i\, \alpha$ is
\bea\label{potax}
&&V_{KL}={1\over 6 t ^2} \Bigl(a A^2 e^{-2 a t } (a t +3)+b B^2 e^{-2 b t } (b t +3)\nonumber\\ &&+3 a A W_{0}\,  e^{-a t }
   \cos  a \alpha  
-3 b B
   W_{0}\,  e^{-b t } \cos b \alpha   \nonumber\\
&&
-A B e^{-(a+b)t} (3(a+b) + 2a  b t )
   \cos    (a-b)\alpha \Bigr)  .
\eea
Despite the fact that we consider a single axion field $\alpha$, we see that the potential has three different periodicities along the axion direction. One may try to describe this property of the potential by introducing effective axion decay constants $f_{a}  \sim 1/at_{0}$,  $f_{b}  \sim 1/bt_{0}$, and   $f_{a-b}  \sim 1/(a-b)t_{0}$.   
 However, this interpretation can be misleading.   

Indeed, nonperturbative terms in the KL superpotential \rf{adssup} are exponentially suppressed for  $f_{a}^{-1} > 1$,  $f_{b}^{-1} > 1$  \rf{KLWGC}. It could be tempting, following  \cite{Moritz:2018sui}, to introduce an additional condition   
\be \label{un}
f_{a-b}^{-1} \sim (a-b) \, t_{0} > 1 \ ,
\ee
similar to the conditions \rf{linear}, or  require that the corresponding instanton-type expansion parameter  is small,
$e^{-(a-b) \, t_{0}} <  1$, as in \rf{large}. 

But it is hard to justify this additional requirement. Indeed, if the expression for the superpotential \rf{adssup} is reliable because of the exponential suppression of the nonperturbative terms, then the expression for the last term in \rf{potax}  is also reliable. No higher order effects suppressed by $e^{-(a-b) \, t_{0}}$ are known to us.  
Thus, in our opinion, the consistency conditions \rf{large} should be sufficient, and until one provides a good reason for an additional constraint \rf{un}, one may simply ignore it.

 However, suppose that one wants to err on the side of caution
and take the speculative condition \rf{un} seriously. As we will see, this is not a problem as well. 
The problem discussed in  \cite{Moritz:2018sui} appears only if one makes two other  steps. First of all, the authors of   \cite{Moritz:2018sui} assumed that $a-b \ll a$, which  is not required in the KL model; for example, it was used in ~\cite{Kallosh:2004yh}, but not in  \cite{Kallosh:2011qk}. Secondly,   the authors of   \cite{Moritz:2018sui}  imposed an additional condition that the last term in \rf{potax} must be smaller than the two previous terms   for $t \sim 1/a \approx 1/b$. 

In our opinion, this last step is unwarranted.  Indeed, in the limit $a-b \ll a$ the conditions \rf{large} imply that $t \sim 1/a \approx 1/b \ll t_{0}$.  Thus the authors of  \cite{Moritz:2018sui} imposed their conditions at $t \ll t_{0}$, at the steep potential wall, very far away from the minimum. This  led them to a strong additional constraint $|W_{0}|\gtrsim \rm{min} (|A|, |B|)$ (eq. (16) of  \cite{Moritz:2018sui}), which played the central role in their investigation, but which is not actually required for the consistency of the KL model. 
 
 The main requirement for any model  of dS vacuum stabilization in string theory is to be valid in the vicinity of the dS minimum of the potential for some range of values $t = {\cal O}(t_{0})$, see e.g. \cite{Baumann:2014nda}. 
 That is exactly  what we did formulating the conditions  \rf{large}, and even the speculative condition \rf{un}. All important features of the KL potential appear at $t \gtrsim t_{0}$.  If the consistency conditions  \rf{large} and   \rf{un} are satisfied at $t \sim t_{0}$, then they are automatically satisfied at all $t \gtrsim t_{0}$. 
  
By using  \rf{sigmacr}, one finds that for $t = {\cal O}(t_{0})$ the coefficient $(a-b)$ in    \rf{un} is canceled by the pole $1/(a-b)$ in \rf{sigmacr}, and the strong form of the weak gravity conjecture
 \rf{un} proposed in   \cite{Moritz:2018sui} is  satisfied for a broad range of parameters such that 
 \be\label{extraWGC}
f_{a-b}^{-1} \sim \ln {a\,A \over b\,B} > 1\ .
\ee

 According to the discussion of the racetrack potential in eq.  (21) of \cite{Burgess:1998jh}, 
\be\label{param}
 \left|{a\,A \over b\,B}\right| =\Big ({\tilde \mu_1\over \tilde \mu_2}\Big )^3 \ .
\ee
 Here  $a = {2\pi/  N_1}$, $b = {2\pi/  N_2}$,  $A =   N_1 \tilde\mu_1^3 $,  $B = N_2 \tilde\mu_2^3$. There are no known restrictions on the size or sign of the ratio $\tilde \mu_2/ \tilde \mu_1$. These quantities depend on the  values of the complex structure moduli, as discussed in  \cite{Baumann:2006th,Baumann:2010sx}. They  may take  different values for each stack of the branes, depending on moduli stabilization. 
 
 To give some particular examples, let us first  consider the case  $b = a/2$. In this case,  one has $ (a-b) \, t =  b \, t$, therefore   the additional constraint \rf{un} introduced in  \cite{Moritz:2018sui} exactly coincides with our second constraint in \rf{linear}. 
Therefore for $b = a/2$ the  weak gravity conjecture  proposed in  \cite{Moritz:2018sui} does not lead to any additional constraints on the parameters of the KL model. One can show that all required conditions, including  $a\,t_0> 1$,  $b\,t_0 > 1$,  and   $(a-b) \, t_{0} > 1$, are satisfied for $A \gtrsim 2B$.  Our results are  consistent with the results of Ref.   \cite{Blanco-Pillado:2018xyn}. 
 Similar results are valid as well for the more general case $a> b \gtrsim a/2$ (including the case $a-b \ll a$) and $A \gtrsim 3B$. We are unaware of any no-go theorems that would forbid having such parameters   in the string theory landscape.

 We conclude,   that for a broad choice   of parameters,  the KL version of the KKLT construction does not violate the standard consistency constraints \rf{large}, and even the   speculative form of the weak gravity conjecture  \rf{un} introduced in \cite{Moritz:2018sui}.

\section{Conclusions}
 In this paper we  note  that the 10d arguments  of \cite{Moritz:2017xto,Gautason:2018gln} against dS uplifting were based on a number of conjectures and an incomplete theory of gaugino condensation, missing important terms      \cite{Hamada:2018qef,Kallosh:2019oxv}. The full 10d theory is expected to provide results compatible with the 4d description of the KKLT scenario. All presently available consistent generalizations of the 4d KKLT model discussed in  \cite{Kallosh:2018wme,Kallosh:2018psh} and in the present paper  reveal the existence of dS vacua in the KKLT scenario.
 
We also argue  that the KL version of the KKLT scenario  \cite{Kallosh:2004yh}, which is based on a racetrack superpotential and does not present any problems with uplifting  \cite{Moritz:2017xto}, is compatible with the weak gravity conjecture for a broad choice of parameters of the KL model.

Thus,  the results of our analysis of all presently available  consistent generalizations of the 4d KKLT  model  do not support the recent swampland conjecture \cite{Obied:2018sgi,Ooguri:2018wrx} and  are fully compatible with the existence of dS vacua in string theory.

\acknowledgments

We would like to thank Sergio Ferrara, Shamit Kachru,  Liam McAllister, Miguel Montero, Jakob Moritz, Susha Parameswaran, Tom Rudelius, Eva Silverstein, Sandip Trivedi, Vincent Van Hemelryck, Thomas Van Riet, Alexander Westphal, and Timm Wrase for helpful comments and discussions. The work  of RK and AL is supported by SITP,  by the NSF Grant PHY-1720397, and by the Simons Foundation grant.  EM is supported in part by the National Science and Engineering Research Council of Canada via a PDF fellowship.  MS is supported by the Research Foundation - Flanders (FWO) and the European Union's Horizon 2020 research and innovation programme under the Marie Sk{\l}odowska-Curie grant agreement No. 665501.

\bibliography{lindekalloshrefs}
\bibliographystyle{utphys}

\end{document}